\documentclass[12pt]{article}%
\usepackage{graphics}%
\def\preprint%
#1{\thispagestyle{empty}~\newline\vspace*{-22.65mm}%
\begin{flushright}%
\begin{tabular}{l} #1%
\end{tabular}%
\end{flushright}%
\vspace{1cm}}%
\renewcommand{\theequation}{\arabic{section}.\arabic{equation}}%
\begin{document}%
\markright{\hfil Deriving relativistic momentum and energy. II}%
\title{\bf \LARGE Deriving relativistic momentum and energy.  II.
Three-dimensional case}%
\author{Sebastiano Sonego\thanks{\tt
sebastiano.sonego@uniud.it}\hspace{2mm} %
and %
Massimo Pin\thanks{\tt pin@fisica.uniud.it}\\[2mm]%
{\small \it  Universit\`a di Udine, Via delle Scienze 208,
33100 Udine, Italy}}%

\date{{\small June 14, 2005; \LaTeX-ed \today }}%
\maketitle%
\begin{abstract}%
We generalise a recent derivation of the relativistic expressions for
momentum and kinetic energy from the one-dimensional to the
three-dimensional case.\\%

\end{abstract}%
\def\p{\mbox{\boldmath $p$}}%
\def\u{\mbox{\boldmath $u$}}%
\def\v{\mbox{\boldmath $v$}}%
\def\w{\mbox{\boldmath $w$}}%
\def\n{\mbox{\boldmath $n$}}%
\def\F{\mbox{\boldmath $F$}}%
\def\bu{\bar{\mbox{\boldmath $u$}}}%
\def\0{\mbox{\boldmath $0$}}%
\def\Cdot{\mbox{\boldmath $\cdot$}}%
\def\Times{\mbox{\boldmath $\times$}}%
\def\Nabla{\mbox{\boldmath $\nabla$}\!}%
\def\Pphi{\mbox{\boldmath $\Phi$}}%
\def\d{{\mathrm d}}%
\def\e{{\mathrm e}}%
\def\SIZE{1.00}%

\section{Introduction}%
\label{sec:intro}%

We have recently shown \cite{sp} how to construct, given a
velocity composition law, the expressions of kinetic energy,
momentum, the Lagrangian and the Hamiltonian for a particle, in a
general mechanical theory satisfying the principle of relativity,
and in which elastic collisions between asymptotically free
particles exist.  For reasons explained in that paper, the
discussion in reference \cite{sp} was restricted to the case of
one spatial dimension only.  In the present article we extend the
treatment to three dimensions.%

Let us briefly review the key elements upon which the derivation
in reference \cite{sp} is based.  The starting point is the
definition of kinetic energy for a particle, as a scalar quantity
whose change equals the work done on the particle. Mathematically,
this amounts to requiring the validity of the fundamental relation%
\begin{equation}%
\d T(\u)=\u\,\Cdot\,\d\p(\u)\;,%
\label{dT}%
\end{equation}%
where $T(\u)$ and $\p(\u)$ are, respectively, the kinetic energy
and the momentum of a particle with velocity $\u$.  For a system
of non-interacting particles, kinetic energy is then necessarily
additive, since work is, so one can easily write down a formula
that expresses energy conservation during an elastic collision. On
requiring that this holds in an arbitrary inertial frame, it
follows that another quantity is conserved, in addition to energy
--- a vector one, that we identify with momentum.  As we shall see,
this quantity is linked to kinetic energy through a simple
equation containing a matrix of functions $\varphi_{ij}$ (the
indices $i$ and $j$ run from 1 to 3), which is uniquely determined
by the composition law for velocities.  This, together with
equation (\ref{dT}) above, allows one to find the explicit
dependence of both energy and momentum on velocity.%

We begin by collecting, in the next section, some results
concerning the velocity composition law in three dimensions.
Section \ref{sec:general} contains the general outline of the
method, which is then applied in section \ref{sec:isotropy} to the
cases of Newtonian and Einstein's dynamics.  A short summary of
the results, together with an assessment about their suitability
from a pedagogical point of view, is given in section
\ref{sec:comments}.  The Appendix contains two mathematical
relations that are not used in the body of the paper, but that
might nevertheless turn out to be useful in further possible
developments.%

\section{Kinematical preliminaries}%
\label{sec:kinematics}%
\setcounter{equation}{0}%

Suppose that a particle moves with velocity $\u$ with respect to a
reference frame $\cal K$.  If $\cal K$ moves with velocity $\v$
with respect to another reference frame $\overline{\cal K}$, the
particle velocity $\bu$ with respect to $\overline{\cal K}$ is
given by some composition law%
\begin{equation}%
\bu=\Pphi(\u,\v)\;.%
\label{comp-gen-3d}%
\end{equation}%
(It is important to appreciate that this relation contains vectors
belonging to two different spaces.  Not only are the basis used to
write the vectors $\v$ and $\bu$, and the vector $\u$, different;
they even span different rest spaces, namely those of
$\overline{\cal K}$ and $\cal K$, respectively.)  The relativity
principle requires that (\ref{comp-gen-3d}) give the composition
law of a group, i.e., that:%
\begin{equation}%
\Pphi(\u,\0)=\Pphi(\0,\u)=\u\;,\;\;\forall\u\;;\\%
\label{id}%
\end{equation}%
\begin{equation}%
\forall\u, \exists \u'\;\;\mbox{such that}
\;\;\Pphi(\u,\u')=\Pphi(\u',\u)=\u\;;\\%
\label{inverse}%
\end{equation}%
\begin{equation}%
\Pphi(\Pphi(\u,\v),\w)=\Pphi(\u,\Pphi(\v,\w))\;,\;\;\forall\u,\v,\w\;.%
\label{ass}%
\end{equation}%
Note that, although in the Galilean case,%
\begin{equation}%
\Pphi(\u,\v)=\u+\v\;,%
\label{gal}%
\end{equation}%
the composition law is commutative, this is {\em not\/} the case
in general, unless velocities are collinear.  For example, the
relativistic law \cite{rindler}%
\begin{equation}%
\Pphi(\u,\v)=
\frac{1}{1+\u\,\Cdot\,\v/c^2}\left[\frac{\u}{\gamma(v)}
+\v\left(\frac{\u\,\Cdot\,\v}{v^2}\left(1-
\frac{1}{\gamma(v)}\right)+1\right)\right]\;,%
\label{Frel}%
\end{equation}%
where, as usual,%
\begin{equation}%
\gamma(v):=\frac{1}{\sqrt{1-v^2/c^2}}%
\label{lorentz}%
\end{equation}%
denotes the Lorentz factor, is not commutative.%

{From} the composition law (\ref{comp-gen-3d}) one can define a
matrix whose components are, in Cartesian bases,%
\begin{equation}%
\varphi_{ij}(\u):=\left.\frac{\partial \Phi_j(\u,\v)}{\partial
v_i}\right|_{\mbox{\footnotesize{$\v=\0$}}}\;.%
\label{fij}%
\end{equation}%
Equation (\ref{inverse}) then imposes that
$\varphi_{ij}(\0)=\delta_{ij}$.  In fact, for the Galilean
composition law (\ref{gal}), we have
$\varphi_{ij}(\u)=\delta_{ij}$ for all $\u$.  On the other hand,
in the relativistic case (\ref{Frel}),%
\begin{equation}%
\varphi_{ij}(\u)=\delta_{ij}-\frac{u_i\,u_j}{c^2}\;.%
\label{fij-rel-3d}%
\end{equation}%
%

\section{General analysis}%
\label{sec:general}%
\setcounter{equation}{0}%

Let $T(\u)$ be the kinetic energy of a particle with velocity $\u$
in an inertial frame $\cal K$. During an elastic collision between
two particles, energy conservation requires that%
\begin{equation}%
T_1({\u}_1)+T_2({\u}_2)=T_1({\u}'_1)+T_2({\u}'_2)\;.%
\label{T+T}%
\end{equation}%
(Of course, the kinetic energy will also depend on the particle
mass; we keep track of this dependence with the indices 1 and 2 on
$T$.)

With respect to another inertial frame $\overline{\cal K}$, in
which $\cal K$ moves with velocity $\v$, the particle velocities
are $\bar{\u}_1=\Pphi(\u_1,\v)$, $\bar{\u}_2=\Pphi(\u_2,\v)$,
$\bar{\u}'_1=\Pphi(\u'_1,\v)$, and $\bar{\u}'_2=\Pphi(\u'_2,\v)$.
Conservation of energy in $\overline{\cal K}$ then implies%
\begin{equation}%
T_1(\bar{\u}_1)+T_2(\bar{\u}_2)=T_1(\bar{\u}'_1)+T_2(\bar{\u}'_2)\;,%
\label{T'+T'}%
\end{equation}%
where we have used the same functions $T_1$ and $T_2$ as in
(\ref{T+T}), because of the relativity principle.%

The expansion of $T(\bu)$ around $\v=\0$ is\footnote{We adopt the
convention of summing over repeated indices.}%
\begin{equation}%
T(\bu)=T(\u)+v_i\,\varphi_{ij}(\u)\,\frac{\partial T(\u)}{\partial
u_j}+{\cal O}(\v^2)\;.%
\label{exp-3d}%
\end{equation}%
Doing this for each term in equation (\ref{T'+T'}), then using
equation (\ref{T+T}) and considering the terms of first order in
$\v$, we find the following additional conservation law:%
\begin{equation}%
\varphi_{ij}(\u_1)\,\frac{\partial T_1(\u_1)}{\partial u_j}
+\varphi_{ij}(\u_2)\,\frac{\partial T_2(\u_2)}{\partial u_j}
=\varphi_{ij}(\u'_1)\,\frac{\partial T_1(\u'_1)}{\partial u_j}
+\varphi_{ij}(\u'_2)\,\frac{\partial T_2(\u'_2)}{\partial u_j}\;.%
\label{p+p}%
\end{equation}%
Thus, one arrives at the conclusion that the vector $\p(\u)$,
whose components are%
\begin{equation}%
p_i(\u)=\varphi_{ij}(\u)\,\frac{\partial T(\u)}{\partial u_j}\;,%
\label{p-3d}%
\end{equation}%
is conserved during a collision, in addition to energy.  In the
one-dimensional case, this quantity can be identified with linear
momentum \cite{sp}, and we suggest doing the same in three
dimensions.\footnote{Actually, the most general conserved quantity
has the form%
\[ \lambda_{ij}\,\varphi_{jk}(\u)\,\frac{\partial T(\u)}{\partial
u_k}+\mu_i\,T(\u)+\nu_i\;, \]%
where $\lambda_{ij}$, $\mu_i$ and $\nu_i$ are a tensor and two
vectors that do not depend on $\u$.  The simplest choice
$\lambda_{ij}=\delta_{ij}$, $\mu_i=\nu_i=0$ corresponds to the
requirement that space be isotropic, and is the one which leads to
the standard expressions for $T(\u)$ and $\p(\u)$.} Note that,
with this definition, linear momentum turns out to be a one-form
rather than a vector, which is very satisfactory from a formal
point of view.%

If we know the function $T(\u)$, we can find $\p$.  If we do not
already know $T(\u)$, we can define it by requiring that it
satisfies the fundamental relation (\ref{dT}), which expresses the
equality between the work done on the particle and the variation
of its kinetic energy \cite{sp}.  On rewriting the differentials
$\d T$ and $\d\p$ in (\ref{dT}) in terms of $\d u_i$ one gets%
\begin{equation}%
\frac{\partial T(\u)}{\partial u_i}=u_j\,\frac{\partial
p_j(\u)}{\partial u_i}\;.%
\label{T/u}%
\end{equation}%
Taken together, equations (\ref{p-3d}) and (\ref{T/u}) allow one
to determine $T(\u)$.  Using again (\ref{p-3d}), one can then find
$\p(\u)$.%

The free particle Lagrangian must satisfy the relation%
\begin{equation}%
p_i(\u)=\frac{\partial L(\u)}{\partial u_i}\;.%
\label{p=dL/du}%
\end{equation}%
Using equation (\ref{p-3d}), we obtain%
\begin{equation}%
\d L(\u)=\varphi_{ij}(\u)\,\frac{\partial T(\u)}{\partial u_j}\,\d
u_i\;.%
\label{dL}%
\end{equation}%
Obviously, it is only for $\varphi_{ij}=\delta_{ij}$ (i.e., in
Newtonian dynamics) that $L=T+\mbox{const}$ --- a feature already
emphasised in reference \cite{sp}.%

Turning now to the Hamiltonian, we need only notice that
(\ref{T/u}) gives, basically, half of Hamilton's equations of
motion for a system with Hamiltonian
$H(\p)=T(\u(\p))+\mbox{const}$.  Indeed,%
\begin{equation}%
u_i=\frac{\partial u_j(\p)}{\partial p_i}\frac{\partial
T(\u)}{\partial u_j}=\frac{\partial H(\p)}{\partial p_i}%
\label{hamilton}%
\end{equation}%
or, symbolically, $\u=\Nabla_p H$.  This allows us to identify
$H(\p)$ with $T(\u(\p))$, up to an additive $\p$-independent
constant.%

\section{Isotropy}%
\label{sec:isotropy}%
\setcounter{equation}{0}%

The previous discussion was general, in the sense that it was
based only on the principle of relativity and on the hypothesis of
space homogeneity (implicit in our use of inertial systems). With
the further requirement that space be isotropic, one can restrict
$\varphi_{ij}(\u)$ to having the functional form%
\begin{equation}%
\varphi_{ij}(\u)=\delta_{ij}+f(u)\,u_i\,u_j\;,%
\label{phi1}%
\end{equation}%
where $f$ is an arbitrary function of the magnitude $u$ of $\u$.
This follows immediately by considering that no other vector
except $\u$ can be used in writing $\varphi_{ij}$.  In fact, even
the class (\ref{phi1}) is too wide, because relativity,
homogeneity, and isotropy together, force $f$ to be a
constant.\footnote{Since $f$ depends only on the magnitude $u$ of
the velocity, this result can be established simply by comparison
with the one-dimensional case \cite{sp,K}.}  Simple physical
considerations \cite{Kgeq0} then require that such a constant be
non-positive, so we shall write from now on%
\begin{equation}%
\varphi_{ij}(\u)=\delta_{ij}-K\,u_i\,u_j\;,%
\label{phi2}%
\end{equation}%
where $K\geq 0$.  The cases $K=0$ and $K=1/c^2$
correspond to the Galilei and Einstein composition law.%

{From} equation (\ref{T/u}) we find%
\begin{equation}%
u_i\,\frac{\partial T}{\partial u_i}=u_i\,\frac{\partial\left(p_j
u_j\right)}{\partial u_i}-p_i u_i\;.%
\label{uT/u}%
\end{equation}%
Inserting (\ref{p-3d}) with the form (\ref{phi2}) for
$\varphi_{ij}$ into (\ref{uT/u}), we obtain%
\begin{equation}%
2\,u_i\,\frac{\partial T}{\partial
u_i}=u_i\,\frac{\partial}{\partial u_i}\left(u_j\,\frac{\partial
T}{\partial u_j}-K\,u^2\,u_j\,\frac{\partial T}{\partial
u_j}\right)+K\,u^2\,u_i\,\frac{\partial T}{\partial u_i}\;.%
\label{Tpart}%
\end{equation}%
Using the mathematical identity%
\begin{equation}%
u_i\,\frac{\partial}{\partial u_i}=2\,u^2\,\frac{\d}{\d
u^2}=2\,\frac{\d}{\d\xi}\;,%
\label{d}%
\end{equation}%
where $\xi:=\ln u^2$, equation (\ref{Tpart}) can be rewritten as%
\begin{equation}%
\left(2+K\,\e^\xi\right)\frac{\d
T}{\d\xi}=2\left(1-K\,\e^\xi\right)\frac{\d^2 T}{\d\xi^2}\;.%
\label{Txi}%
\end{equation}%
This is a simple differential equation for $T$, that can be
solved by first finding $\d T/\d\xi$:%
\begin{equation}%
\frac{\d T}{\d\xi}=A\,\exp\left(\frac{1}{2}\int\d\xi\,\frac{2
+K\,\e^\xi}{1-K\,\e^\xi}\right)
=\frac{A\,\e^\xi}{\left(1-K\,\e^\xi\right)^{3/2}}\;,%
\label{T/xi}%
\end{equation}%
where $A$ is an arbitrary constant.  By a further integration we
find, for $K\neq 0$,%
\begin{equation}%
T(\u)=\frac{A/K}{\left(1-K\,u^2\right)^{1/2}}+B\;;%
\label{TKneq0}%
\end{equation}%
while, for $K=0$,%
\begin{equation}%
T(\u)=A\,u^2+B\;.%
\label{TK=0}%
\end{equation}%
In both cases, $B$ is a further integration constant.  On
requiring that $T({\bf 0})=0$, we can write $B$ in terms of $A$,
which can then be expressed using the more familiar parameter $m$
--- the particle mass.  We thus obtain, for $K\neq 0$,%
\begin{equation}%
T(\u)=\frac{m/K}{\left(1-K\,u^2\right)^{1/2}}-m/K\;;%
\label{T1}%
\end{equation}%
and, for $K=0$,%
\begin{equation}%
T(\u)=\frac{1}{2}\,m\,u^2\;.%
\label{T2}%
\end{equation}%
Note that (\ref{T2}) coincides with the limit of (\ref{T1})
for $K\to 0$.  For $K=1/c^2$ one recovers the standard expression
for kinetic energy in Einstein's dynamics,%
\begin{equation}%
T(\u)=\frac{m\,c^2}{\left(1-u^2/c^2\right)^{1/2}}-m\,c^2\;.%
\label{TE}%
\end{equation}%
It is then a straightforward exercise to find the usual
expressions for momentum, the Lagrangian, and the Hamiltonian,
using equations (\ref{p-3d}), (\ref{dL}), and the remarks at the
end of section \ref{sec:general}.  The results are the same as
in the one-dimensional case, so we do not list them here.%

In closing this section, we note that the calculations can be
simplified by arguing that $T(\u)$ can only be a function of $u$,
and that $\p(\u)$ must be of the form $\p(\u)=\alpha(u)\u$, with
$\alpha$ an unspecified function of $u$.  However, we have
preferred not to rely on these results, because they need not be
derived separately, but are actually consequences of equations
(\ref{p-3d}), (\ref{T/u}), and (\ref{phi2}).%

\section{Conclusions}%
\label{sec:comments}%
\setcounter{equation}{0}%

In this paper we have extended the derivation of relativistic
energy and momentum given in \cite{sp}, from one to three
dimensions.  Although nothing changes in the results, it is
obvious that the discussion is not as elementary as for the
one-dimensional case, since it requires some familiarity with
multivariable calculus.  Hence, this material is not suitable for
an introductory course, contrary to what happens for the treatment
in reference \cite{sp}.  It can, however, be presented in a
standard course on analytical mechanics, at the level, e.g., of
reference \cite{goldstein}.  Indeed, we believe that students
would benefit from being exposed to this approach, which relies on
a very small number of physical hypotheses, and has therefore the
advantage of being logically very simple.%

\section*{Appendix: Differential constraints}%
\setcounter{equation}{0}%
\renewcommand{\theequation}{A.\arabic{equation}}%

We present two mathematical relations --- equations
(\ref{condition}) and (\ref{psi-diff-disp-3d}) below --- that have
not been used in the body of the article, but that nevertheless
are somewhat interesting by their own, since they represent
differential constraints on momentum and on the Hamiltonian.%

Replacing (\ref{T/u}) into (\ref{p-3d}), one arrives at a single
equation for $\p(\u)$:%
\begin{equation}%
p_i(\u)=\varphi_{ij}(\u)\,\frac{\partial p_k(\u)}{\partial u_j}\,u_k\;.%
\label{condition}%
\end{equation}%
On defining $\psi_{ij}(\u)$ as the components of the inverse
matrix of the $\varphi_{ij}(\u)$, that is
$\psi_{ij}\,\varphi_{jk}=\delta_{ik}$, equation (\ref{condition})
can be rewritten as%
\begin{equation}%
\frac{\partial p_j(\u)}{\partial u_i}\,u_j=\psi_{ij}(\u)\,p_j\;.%
\label{newcond}%
\end{equation}%
Taking now the second derivative of Hamilton's equations
(\ref{hamilton}) we get%
\begin{equation}%
\frac{\partial u_i}{\partial p_j}=\frac{\partial^2 H}{\partial
p_i\,\partial p_j}\;.%
\label{du}%
\end{equation}%
Replacing (\ref{du}) into equation (\ref{newcond}), we arrive at
the following differential constraint on $H(\p)$:%
\begin{equation}%
\frac{\partial H}{\partial p_i}=\frac{\partial^2 H}{\partial
p_i\,\partial p_j}\,\psi_{jk}(\Nabla_p H)\,p_k\;.%
\label{psi-diff-disp-3d}%
\end{equation}%
%

%
\end{document}